\documentclass[print]{revtex4-1}
\pdfoutput=1
\usepackage[lofdepth,lotdepth]{subfig}
\usepackage{graphicx}
\usepackage{natbib}
\usepackage{float}
\usepackage{amsmath}
\usepackage{amssymb}
\usepackage{epsfig}
\usepackage{subeqnarray}
\usepackage{color}
\usepackage{bbold}
\usepackage{dsfont}
\usepackage{tikz}
\usepackage{cancel}
\usetikzlibrary{positioning,calc}
\usepackage{hyperref}

\begin{document}

\title{Expanding materials selection via transfer learning for high-temperature oxide selection}

\author{Zachary D. McClure}

\affiliation{School of Materials Engineering and Birck Nanotechnology Center, Purdue University, West Lafayette, Indiana 47907, USA}
\author{Alejandro H. Strachan}
\affiliation{School of Materials Engineering and Birck Nanotechnology Center, Purdue University, West Lafayette, Indiana 47907, USA}

\affiliation{Corresponding author: strachan@purdue.edu}

\date{\today}
\begin{abstract}
Materials with higher operating temperatures than today's state of the art can improve  system performance in several applications and enable new technologies. Under most scenarios, a protective oxide scale with high melting temperatures and thermodynamic stability as well as low ionic diffusivity is required. Thus, the design of high-temperature systems would benefit from knowledge of these properties and related ones for all known oxides. While some properties of interest are known for many oxides (e.g. elastic constants exist for over 1,000 oxides), melting temperature is known for a relatively small subset. The determination of melting temperatures is time consuming and costly, both experimentally and computationally, thus we use data science tools to develop predictive models from the existing data. The relatively small number of available melting temperature values precludes the use of standard tools; therefore, we use a multi-step approach based on sequential learning where surrogate data from first principles calculations is leveraged to develop models using small datasets. We use these models to predict the desired properties for nearly 11,000 oxides and quantify uncertainties in the space.

\end{abstract}

\maketitle

\section{Introduction}
Materials capable of operating at high temperatures are critical for applications ranging from aerospace to energy \cite{pollock2006nickel} and increasing their operating envelope over the current state of the art is highly desirable. For example, increasing the operating temperature of land-based turbines by 30 $^{\circ}$C would result in an approximate 1\% efficiency increase and can translate into sector-wide fuel savings of \$66 billion with significant environmental impact over a 15 year period \cite{Electric2018}. In addition, high temperature metallic alloys can enable rotation detonation engines for hypersonic vehicles \cite{falempin2006toward}. In all of these applications, high-temperature mechanical integrity or high strength are required, and so is oxidation resistance. The latter can be achieved either by the formation of a protective oxide scale during operation \cite{doychak1991protective} or by the incorporation of a protective oxide (often sacrificial) during fabrication \cite{justin2011ultra,smialek2014oxidation}. This article combines existing experimental and first principles data with data science tools, including uncertainty quantification, to create a comprehensive dataset of potential oxides and the physical properties relevant for materials selection. 

In recent years, complex concentrated alloys (CCAs, multi principal component alloys that lack a single dominant component) and the closely related high-entropy alloys (HEAs) \cite{cantor2004microstructural,yeh2004nanostructured} have attracted significant attention as they have been shown to exhibit properties not possible with traditional metallic alloys \cite{gorsse2018high}. Particularly interesting for high-temperature applications are refractory CCAs (RCCAs) which \cite{senkov2010refractory} have emerged as an attractive alternative to current superalloys. While they exhibit high-temperature strength surpassing the state of the art, their oxidation resistance is far from ideal. For example, the mass gain at T=1000$^{\circ}$C for TiZrNbHfTa during 1 hour in air is 65 mg/cm$^{2}$, almost an order of magnitude higher than the Cr$_{2}$O$_{3}$-forming wrought Ni-based superalloys \cite{chang2018oxidation,smith1990high}. Thus, efforts are underway to design RCCAs capable of growing effective oxide scales at temperatures above 1000$^{\circ}$C  \cite{butler2017high,chang2018oxidation}. Beyond RCCAs high-temperature protective oxides are required in a range of applications. Carefully engineered oxide scales can be used to prevent further oxidation and embrittlement of alloys in high temperature applications \cite{smith1990high,chang2018oxidation}, corrosion resistance in adverse environments \cite{smith1990high}, or as a protective coating during aerospace re-entry applications \cite{justin2011ultra}.

Desirable properties in these oxides include high-melting temperatures, good thermodynamic and mechanical stability to facilitate their formation over competing oxides, and low oxygen ion and cation mobility to slow down oxidation kinetics. Other properties are also desirable: a coefficient of thermal expansion (CTE) matching that of the substrate, a Pilling-Bedworth ratio (defined as $\frac{V_{oxide}}{V_{metal}}$ with V the molar volume) near one, and good adhesion to the substrate \cite{bedworth1923oxidation}. Designing RCCAs with desirable oxide scales presents additional challenges since the large number of metallic elements results in various possible, competing, oxides and complex multi-layer scales \cite{chang2018oxidation}. The design of RCCAs with appropriate high-temperature oxidation resistance and the selection of oxide coatings that can be added to structures would benefit enormously from an extensive database of all possible high-temperature oxides and their properties of interest. 

Unfortunately, the required information is not available for the majority of the tens of thousands of stable oxides known. To date, over 60,000 metastable oxides have been studied by the Materials Project (MP) via first principles calculations \cite{Jain2013}. Of these, about 11,000 are either the ground state or low-energy metastable structures at zero temperature. Elastic constants are known for a small sub-set, totaling roughly 1,000 oxides in the MP database \cite{Jain2013}. However, melting temperatures are known for an even smaller subset. In this paper we use data science tools including machine learning to generate materials property information that can used for materials selection for the majority of known oxides. We build on the fact that some of these properties are correlated to each other due to similar underlying physics to address the challenge of small data sets.
  
\textbf{Cyber-infrastructure for materials data}
Motivated by the need for faster and less expensive materials discovery and deployment cycles \cite{national2011materials},  great strides have been made in the development of cyberinfrastructure for materials science and engineering over the last decade. Examples of this infrastructure include open and queryable repositories with first principles data, such as MP or the Open Quantum Materials Database (OQMD) \cite{Jain2013,saal2013materials}, open repositories of materials properties such as Citrination \cite{Informatics2020}, and even published interatomic potential models for atomistic simulations \cite{OpenKIM}. In addition to data and models, platforms for online simulations and data analysis such as nanoHUB \cite{strachan2010cyber} and Google Colab \cite{google} lower the barrier of access to simulation and data science tools for research and education \cite{reeve2019online}.
These repositories are making strides towards making data findable, accessible, interoperable, and reproducible (FAIR) \cite{wilkinson2016fair}. Data can be queried through online user interfaces or via application programming interfaces (APIs) for rapid querying and analysis of data. 

\textbf{Transfer learning for materials selection.}
Materials selection requires access to data and often involves a multi-objective optimization \cite{ashby1993materials,ashby2000multi}. This was traditionally done with existing experimental data, sometimes combined with simple models \cite{ashby1993criteria}. More recently, \textit{ab initio} electronic structure calculations have been incorporated to such efforts \cite{cutello2005class} and progress in multiscale modeling is providing additional tools to materials design and optimization \cite{van2020roadmap}. In addition to such data, machine learning tools are being used to assess the current state of knowledge and make decisions. In our application, it would be tempting to use machine learning to develop models to predict our quantities of interest (QoI), such as melting temperature, from composition using the available data; these models could then be used to explore the properties of a wide range of oxides. Unfortunately, the limited set of known melting temperatures precludes such an approach as standard ML methods require vast amounts of data. The lack of data is common in materials applications and several approaches have been developed to address it \cite{cubuk2019screening}. These methods compensate the lack of large amounts of data with domain expertise, physics, and chemistry. One such method is to enhance the information fed to the model by adding surrogate properties as inputs. These surrogate properties should be both easy to obtain and be expected to correlate to the quantity of interest. In this paper, we use the oxide stiffness (easily computable via \textit{ab initio} simulations) as an additional input to the model to predict melting temperatures. Since both stiffness and melting temperature are governed by the strength of the inter-atomic interactions, there is a correlation between these properties and adding stiffness as an input to the models results better accuracy.

\section{Currently available data}\label{sec:existingdata}

The design or selection of protective oxide scales would benefit from access to materials properties for all possible oxides that are either stable or metastable at the operating temperatures. As discussed above, a large number of oxides structures are known, but high-temperature data, including melting temperatures are known for a small subset. Thus, we start from all known oxides and combine existing data with machine learning to provide information about structures for which we lack experimental data. This section explores the relevant data available in online repositories and Sec. \ref{seqlearn} discusses the use, combination, and extension of the data. 

As discussed above, several materials data repositories focusing on various types of data and materials classes are available today. We leveraged the MP database, Citrination, and WolframAlpha \cite{Wolfram}. The MP is a database with density functional theory (DFT) results including crystal structure data, relative stability to the ground state, elastic constants for select materials, calculated X-ray diffraction (XRD) and X-ray photoelectron spectoscopy (XPS) spectra, and even T=0 K phase diagrams for compounds. MP has information about a majority of known oxides and we start our search within this list. The properties in the MP can be accurately calculated from first principles calculations; however, properties like melting temperature are computationally too intensive for high-throughput DFT calculations. Therefore we turn to repositories with extended datasets for additional information like melting temperature, CTE, and oxygen vacancy formation energy (VFE). 

The Citrination database \cite{Informatics2020} is an open repository where researchers can upload their own data and share it with the community at large. At the time of writing, citrination contains 454 public databases curated by public users and Citrine staff. Databases previously curated through research efforts have been published in their database and are freely available for download and use. For our efforts we turned to Citrination for databases of CTE and oxygen VFE. 

We were unable to find an electronic database with melting temperatures for oxides, most reported melting points exist within individual papers, collected handbooks, or commercial databases. However, we were able to find some of these properties in WolframAlpha, a general purpose, queryable, compute engine. Using WolframAlpha, we generated a list of melting temperatures for a subset of the oxides queried from MP with elasticity data. 

\subsection{The Materials Project: basic oxide data}
\begin{figure}
\includegraphics[scale=0.50]{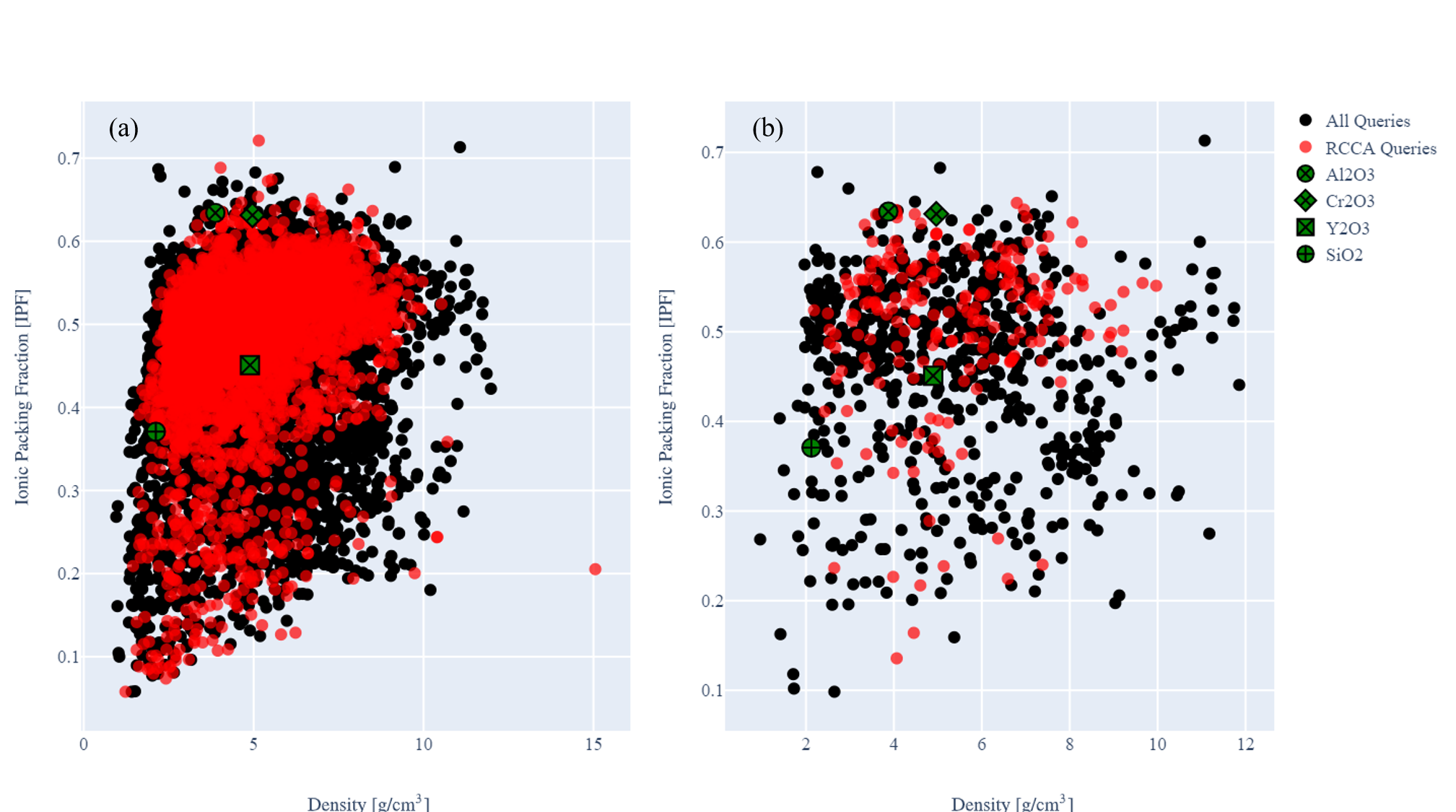}
\caption{Calculated ionic packing fraction of oxides and their queried densities. a) Database curated post energy stability filtering. b) Database curated post elasticity filtering}
\label{fig:IPF_vs_density}
\end{figure}

We accessed the MP database using the Pymatgen API \cite{Ong_2015} and analyzed every oxide available. MP contains information about 60,000 distinct oxides (differing either by composition or crystal structure). All these structures were obtained by energy minimization using DFT within the generalized gradient approximation and additional details of the calculations can be found in work by Jain et al. \cite{jain2011high}. Confirming the metastability of these structures would require positive phonon frequencies and elastic constants to discard local energy maxima; these quantities are not available for all these oxides. To address this challenge we first filtered the data to retain only structures that are 1 meV eV above the convex hull (i.e. the predicted ground state for that composition). We note that the energies resulting from energy minimization correspond to a temperature of T=0 K (minus zero point energy) and phases with free energy higher that the ground state at 0 K can be stabilized at higher temperatures due to entropic contributions to the free energy. Furthermore, many metastable structures are long-lived and used in applications. After this stability constraint, we are left with $\sim$11,000 possible oxides. However, elastic constants are documented in MP (from DFT calculations) for a subset of 855 oxides. To illustrate the available data, Figure \ref{fig:IPF_vs_density} compares two properties of the available oxides after filtering by energy stability and elastic constants. We plot the ionic packing fraction (defined as the total atomic volume assuming hard spheres with the corresponding ionic radius in the unit cell divided by the cell volume) vs. density obtained from the crystal structure data. Red points indicate oxides with at least one element that is found in RCCAs: we select Ti, V, Cr, Zr, Nb, Mo, Ru, Hf, Ta, W, and Re as well as Al, Cr, and Si since they are useful additives. Of the 855 oxides with elasticity data, 235 of them contain an element pertaining to an RCCA or additive compound. The figure also highlights common protective oxides Al$_{2}$O$_{3}$ and Cr$_{2}$O$_{3}$; as expected these oxides have high packing fractions (which correlates in low ionic diffusivity). Interestingly, there are a number of potential compounds with comparable properties to both these common oxides.

With this basic information at hand, we now focus on the remaining properties: melting temperature, ion mobility, and CTE. While these properties can be obtained, in theory,  from first principles they are computationally very intensive and they are not included in the MP. Therefore we turn to other repositories for additional DFT and experimental data.

\subsection{WolframAlpha: melting temperatures}

\begin{figure}[H]
\includegraphics[scale=0.50]{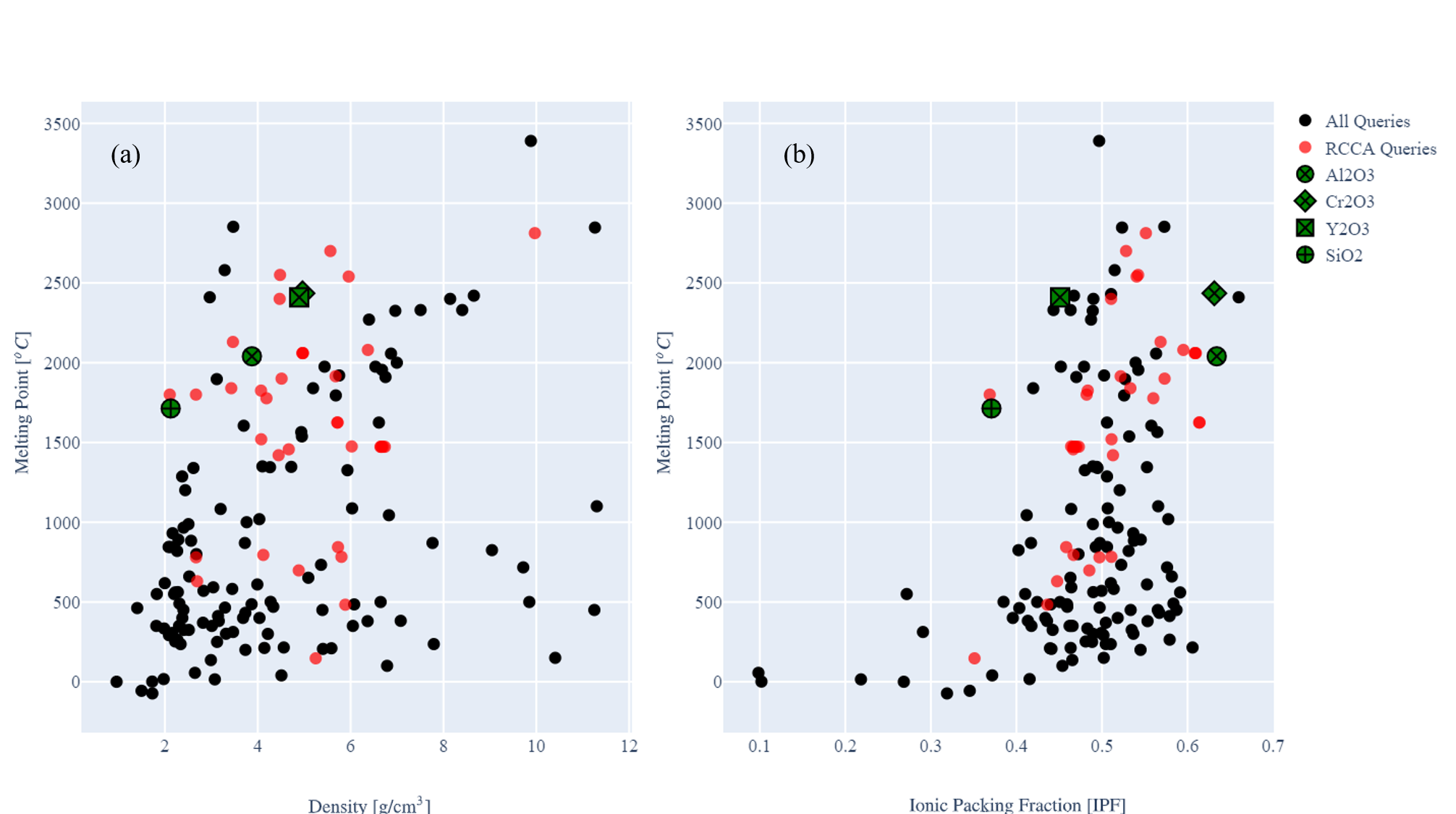}
\caption{Queried melting points from WolframAlpha with bulk modulus (a) and IPF (b) properties.}
\label{fig:temp_melting}
\end{figure}

At the time of writing, melting temperatures of the oxides of interest were not available in materials-specific online repositories. A single curated inorganic melting point database on Citrination exists, but many of the values are not oxides, and do not overlap where we have existing elasticity data. Fortunately, WolframAlpha provides an API for data exploration. Through a series of string queries we obtained and curated melting points of 158 oxides into our database. Since this data is significantly less abundant than the elasticity data from Materials Project we will consider the melting point to be our harder to acquire, or more expensive, set of data. Improvements to this dataset can be made through literature searches, or analyzing phase diagram textbooks, but our goal is to illustrate a rapid acquisition of data rather than the traditional task of searching through physical copies of information. Figure \ref{fig:temp_melting} shows the results of the melting point query with respect to density and IPF properties. RCCA contaning oxides are highlighted to guide the eye, and as expected a number of them have comparable properties to common oxides such as Al$_{2}$O$_{3}$ and Cr$_{2}$O$_{3}$.

\subsection{Citrination: vacancy formation energies}
Ionic mobility is another critical material property in the design of protective oxides, unfortunately, ionic mobility (oxygen or cation) is not widely available. However, since oxygen mobility is mediated by vacancies, the vacancy formation energy is a good surrogate for ionic transport: the higher the vacancy formation, the lower the vacancy concentration and oxygen ion mobility. Citrination includes a database of nearly 2,000  charge neutral vacancy formation energies of oxides based on first principles approaches originally published by Deml et al. \cite{deml2015intrinsic}. Of this dataset 1,200 were unique oxide compounds. 

Another database of importance is available based on work from by Shick et al. \cite{schick2017descriptors}, it contains 69 average CTE values obtained from anharmonic phonon calculations. We will consider the use of this database in future work, but at this time the limited dataset provides a great challenge to accurate predictions outside the selected compounds via machine learning methods. 

\section{Extending oxide data via data-driven sequential learning} \label{seqlearn}

In summary, from online accessible databases we were able to extract 11,000 stable and metastable oxides from an initial 60,000 query on MP. Of these 11,000 possible oxides roughly 1,000 have existing elasticity data. From the list of 1,000 oxides with elasticity information only 162 melting points were obtained through queries. In addition, we have VFE values for 1,200 cases and CTE for 69. Ultimately our goal is be to build models with each of these properties, and use the information leveraged from each to extend a materials search into the original 11,000 stable and metastable oxides. 

Since filling in the gaps in data discussed in Section \ref{sec:existingdata} via first principles calculations or experiments would be prohibitive in terms of time and cost, we will explore using data-driven machine learning tools like neural networks \cite{coley2019graph,elton2018applying} and random forests \cite{ling2017high}. One could attempt to train models that relate composition to the final QoI (e.g. melting temperatures) from the existing data. However, standard ML approaches are not applicable directly due to the scarcity of the data. This is a common challenge in the field of materials. Feature engineering, which involves feeding additional data to the model that can be easily obtained from the raw input data, can be used to address this challenge. For example, we could include electronegativity and ionic radii of the elements as inputs to the model, thus, we include information about bonding and packing. In addition to such {\it periodic table} data, one can further increase the information fed into the model by adding a material property that is easier to obtain than the QoI and that is expected to correlate with it. It has been shown that even with limited training data physics-based descriptors have had a significantly higher impact than models that only rely on raw volumes of data \cite{cubuk2019screening,raissi2019physics}. Here we will build on the fact the melting temperature is governed by similar physics to another property that is more widely known: stiffness. In addition to stiffness we will explore other composition based descriptors with physics built into them.

\subsection{Descriptors} \label{descriptors}

As mentioned above, a common way of building physics into ML models is to use  periodic table data of the elements involved as inputs. We primarily use the {\it composition featurizer} from Matminer \cite{ward2018matminer} to generate a variety of properties with composition as the only input. As shown in previous work by Ward et al. \cite{ward2016general}, statistical descriptors based on the chemical formula are useful for machine learning features.

The descriptors we use are described as the follows:
\begin{enumerate}
  \item A stoichiometric calculation of fractions of elements without considering the actual composition. This calculation includes number of elements in the compound and normalizations of the respective fractions.
  \item Periodic table type descriptors including mean, mean absolute deviation, range, minimum, maximum, and mode of elemental properties. These values include maximum row on the periodic table, average atomic number, and the difference in atomic radii in all elements present.
  \item As previously shown by Meredig et al. \cite{meredig2014combinatorial}, electronic structure attributes with averages of s, p, d, and f valence shell electron concentrations are useful as descriptor inputs.
  \item Assuming that the ionic species in the oxide can form a single oxidation state, an adaption of the fractional ionic character of a compound can be used based on an electronegativity-based measure \cite{callister2018materials}.
  \item The fraction of the transition metal elements.
  \item The cohesive energy per atom using elemental cohesive energies.
  \item An estimation of the band gap center based on electronegativity.
  \item Number of available oxidation states in the compound.
  \item For mechanical properties we also extend descriptors to include properties queried from the MP database like density, space group number, and calculated ionic packing fractions.
  \item For the melting point of a material we will transfer our predicted stiffness properties information to build further descriptors.
\end{enumerate}

These descriptors are able to characterize the output properties of CTE, and VFE sufficiently, and we do not see evidence of over parameterization of the models. For stiffness we add additional descriptors queried from MP, and for the melting point we use the full knowledge of composition descriptors, queried MP properties, and predicted stiffness. 

\subsection{Predictive models for melting temperature using random forests}

Random forests (RFs) approach regression methods through a series of decision trees \cite{ho1998random,breiman2001random} whose outputs are averaged. This averaging is done to overcome the limitation of individual tree predictions which may have difficulty assessing noise or non-linearities in the data. Importantly, progress has been made in the quantification of uncertainties in RFs by Efron \cite{efron2012model} and Wager et al. \cite{wager2014confidence}, and more recently by Ling et al. \cite{ling2017high} with the addition of an explicit bias term to the uncertainty. Neural networks, often outperforming random forest predictions, were considered for this study, but quantification of uncertainty in their outputs is still an active field of research \cite{tripathy2018deep}. Due to the accessibility of uncertainty quantification we choose to implement random forest models with the state of the art uncertainty calibration proposed by Ling et al \cite{ling2017high}. It involves sample-wise variance defined as the average of the jackknife-after-bootstrap and infinitesimal jackknife variance estimates with a Monte Carlo sampling correction. The RF models implemented in this study are available in the Lolo scala library \cite{hutchinson2016citrine}.

We use 350 trees with an unrestricted maximum depth for our RFs. The maximum depth parameter cutoff is defined by the nodes increasing until the leaves become pure, or until the all leaves contain less than two samples. This is the default parameter for Lolopy. Beyond 350 trees the loss function saturates and no further improvement was detected. This is consistent with previous work in the literature \cite{oshiro2012many}.

As is common practice, each descriptor descriptor is normalized by standard normalization and data was split into 80\% training and 20\% testing to evaluate performance. Assessment of the model was performed for each material property by reshuffling the dataset 10 different times, and taking an aggregated MAE.

When assessing uncertainty estimates for an individual output $x$, the residuals, $r(x)$, of the prediction when normalized by the uncertainty $\sigma(x)$ ($N = \frac{r(x)}{\sigma(x)}$), should have a Gaussian distribution with zero mean and unit standard deviation. This metric can help quantify if the random forest uncertainty predictions are well calibrated with respect to the inherent error predicted. 

Using the set of descriptors and architecture detailed in Sec. \ref{descriptors}, we implement random forest models to predict the set of desired properties using databases from MP, WolframAlpha, and Citrination. All reported MAE values are taken as an aggregate mean after shuffling the training and testing sets 10 times.

\textbf{Random forest performance for VFE}

Using the curated Citrination dataset we developed a RF model for VFE. Composition based descriptors obtained via Matminer were used for model predictions. For 10 shuffling samples we report an aggregated MAE of 0.17 eV/atom. 

\begin{figure}[H]
\centering
\includegraphics[scale=0.50]{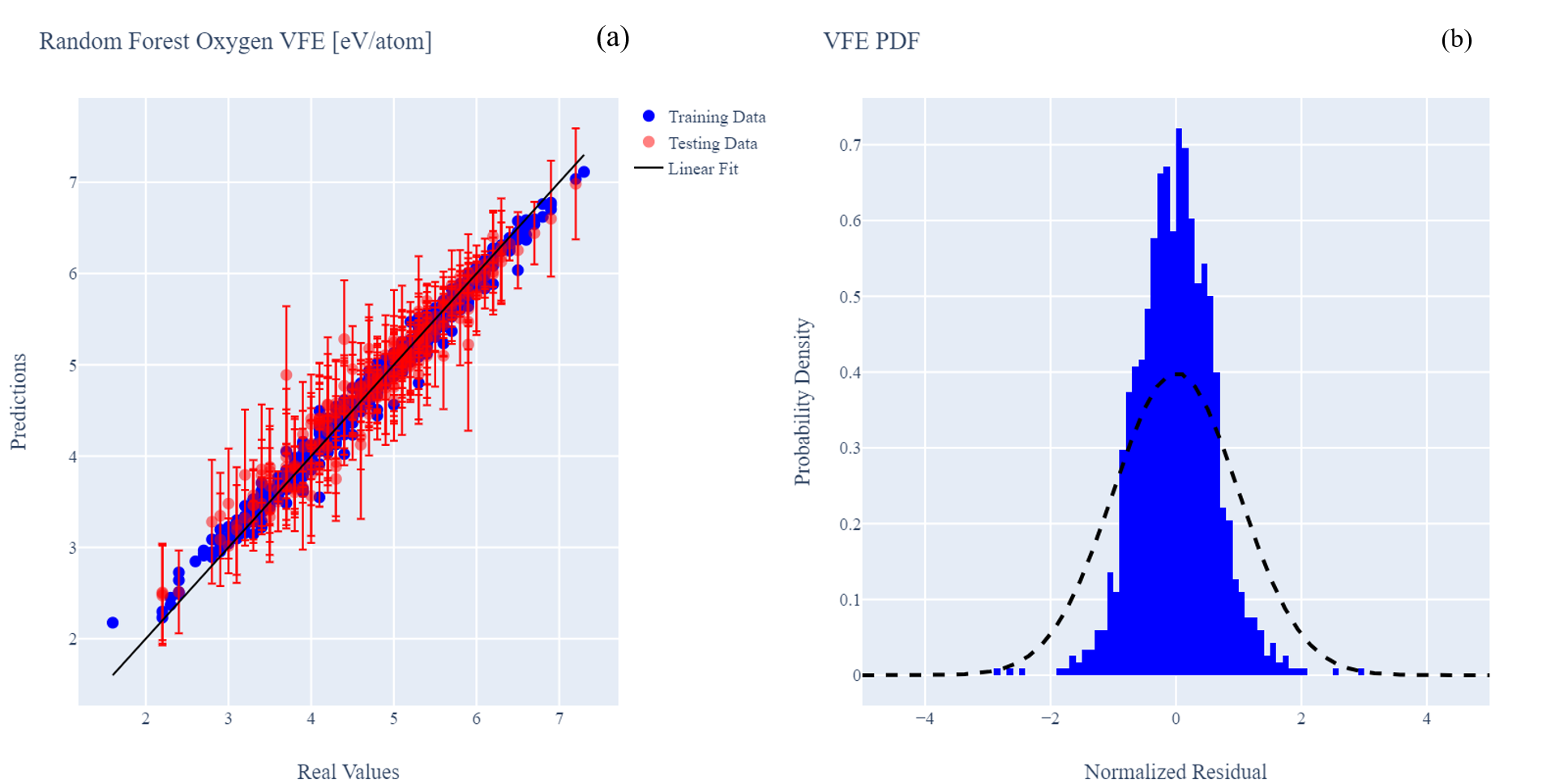}
\caption{a) Parity plot diagram for predicted and real values of oxide VFE. Values directly on the line are a perfect match. b) Normalized residuals for VFE with Gaussian like distribution.}
\label{fig:VFE_RF}
\end{figure}

\textbf{Random forest performance for stiffness}

In addition to the Matiminer featurizers described above we add additional descriptors such as IPF and space group number since these were easily queried. An aggregated testing MAE score of 18 and 10 GPa for bulk and shear modulus was reported after 10 shuffling of samples.

\begin{figure}[H]
\centering
\includegraphics[scale=0.75]{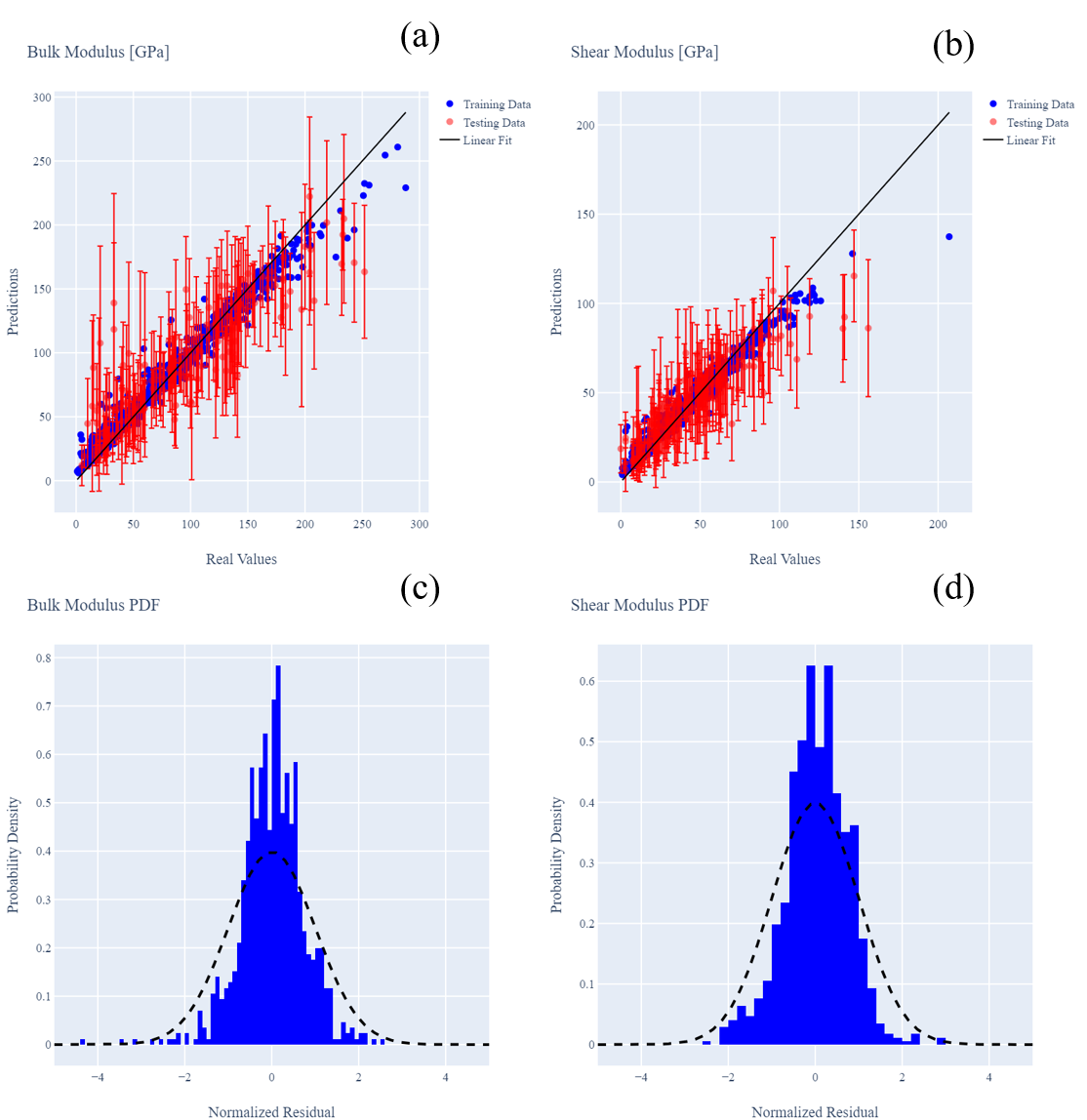}
\caption{a) Parity plot diagram for predicted and real values of oxide bulk modulus, b) shear modulus. Included are normalized residual calculations for c) bulk modulus, and d) shear modulus.}
\label{fig:mech_prop_RF}
\end{figure}

\textbf{Random forest performance for melting temperature}

\begin{figure}[H]
\centering
\includegraphics[scale=0.75]{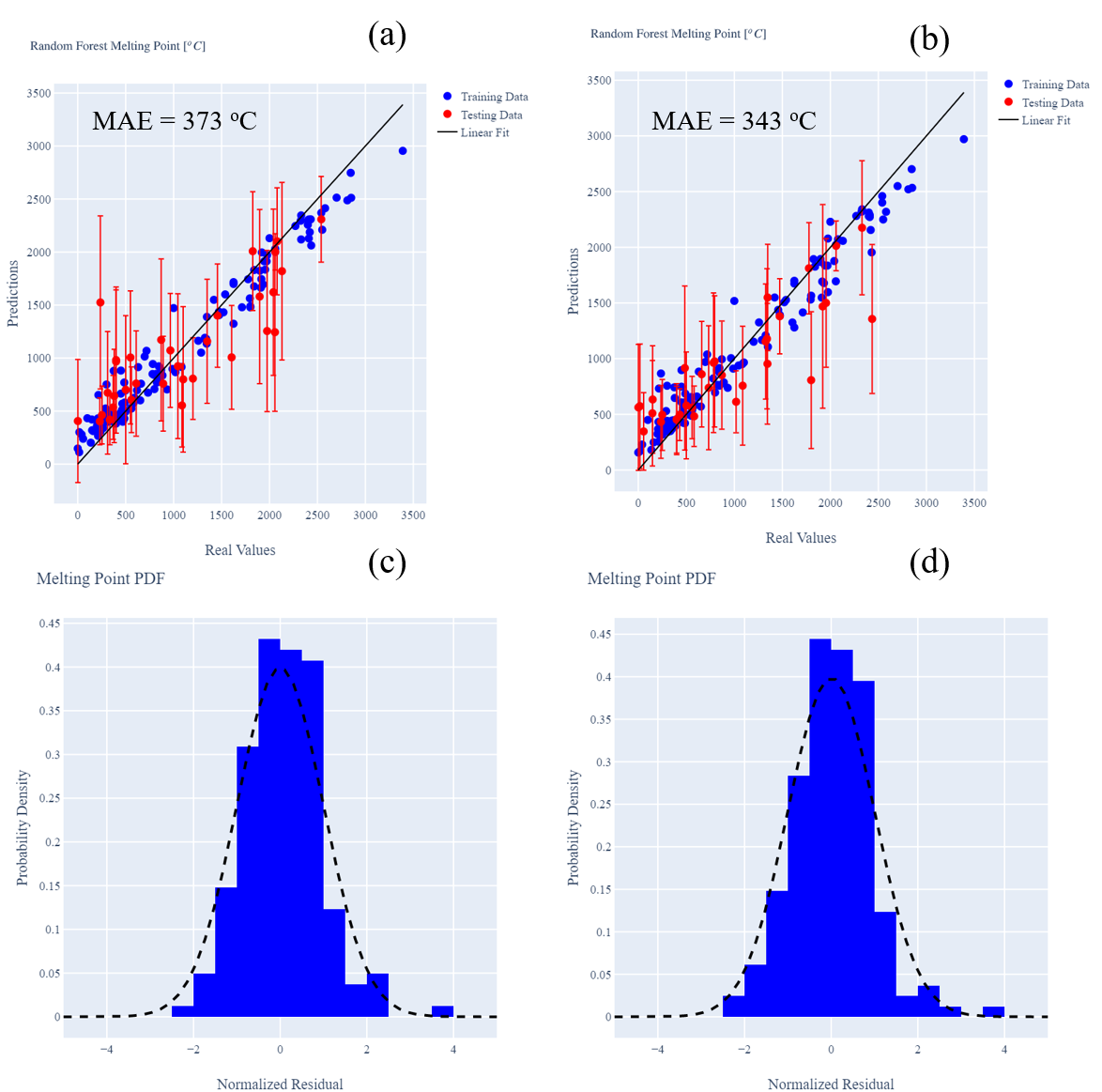}
\caption{a) Parity plot diagram for predicted and real values of oxide melting temperature. Values directly on the line are a perfect match. b) Adding stiffness properties to the model causes a decrease of 40 $^{o}$C with respect to MAE and a noticeable decrease in uncertainty.}
\label{fig:T_melt_RF}
\end{figure}

Our dataset of 162 melting points with corresponding elasticity data was used to create a predictive model for varied oxides. Fig. \ref{fig:T_melt_RF} shows the performance of both the training and testing data before and after adding stiffness properties into the model. As we can see adding stiffness as a descriptor improves the accuracy of the model to a marginal degree with a reduction of the MAE from 373 to 343 $^{o}$C. While we do see a large prediction in the uncertainty, these values are promising for an initial sweep of potential oxides. A noticeable reduction in uncertainty can be seen between the two figures, and after adding stiffness fewer points lie outside the linear fit in the parity plot. After training the model we use identical descriptors for the remaining compounds that we were unable to easily obtain melting points for and extend our predictions using the information gained from stiffness and melting point models. In Sec. \ref{UQ} we will assess some the sensitivity of this prediction with varied UQ methods. In the outlook section of this paper we will discuss the implications and results of extrapolating our predictions to other oxide melting points outside of our initial query with WolframAlpha.

\section{Materials selection for protective oxide scales} \label{selection}

Using the models above we begin to extend our search space of potential oxides from our initial query of 162 melting points and 855 points with elasticity data and move into the space of the remaining 11,000 stable oxides from MP. First, we predict the elasticity data of the remaining 11,000 oxides that did not have this data to begin with. Then we use those descriptors to expand our melting point database from 158 queried data points to nearly 11,000 data points: a two order of magnitude increase.

\begin{figure*}
\includegraphics[scale=0.55]{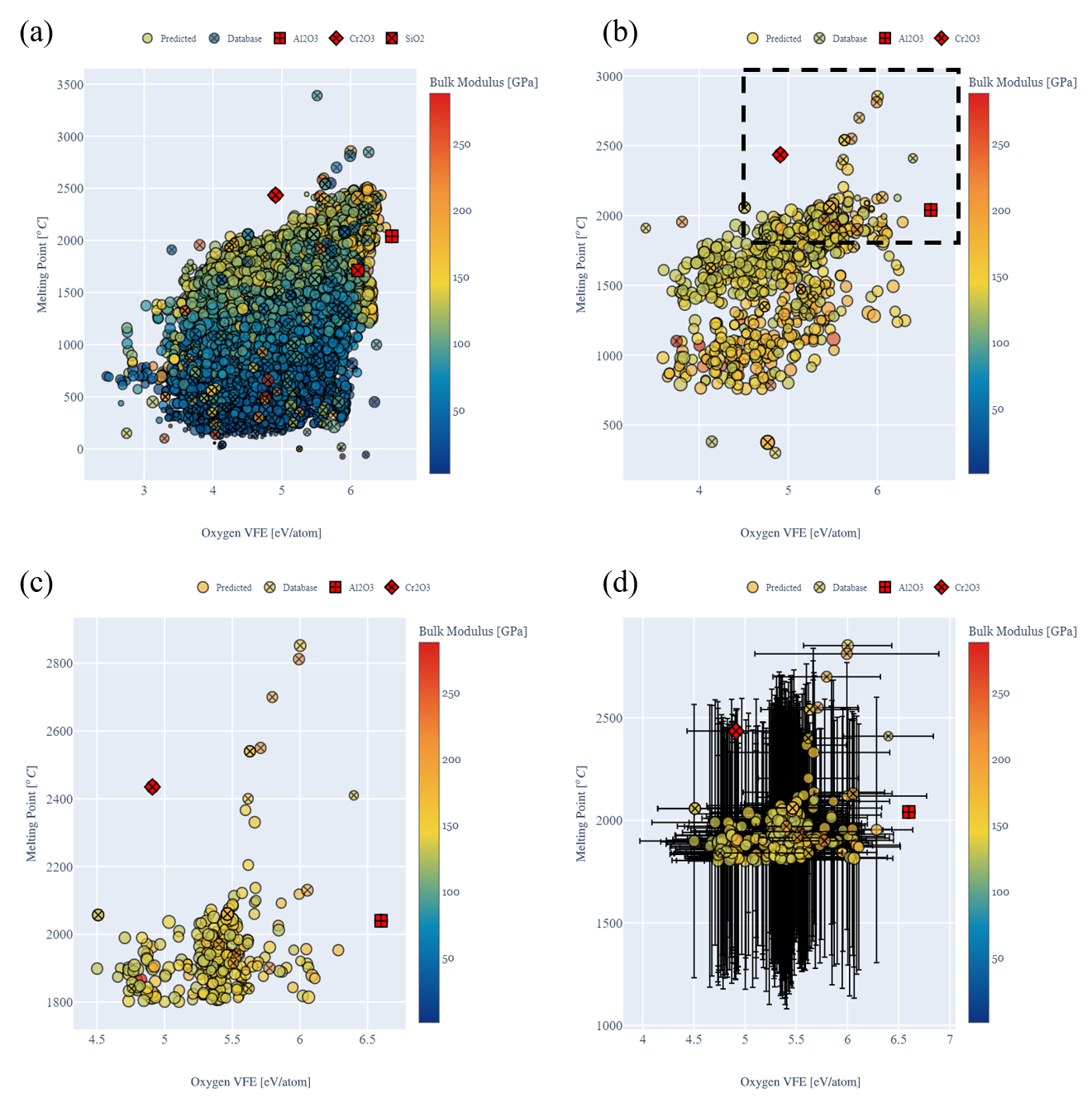}
\caption{Comparison of melting point and vacancy formation energy of oxide compounds. Coloring corresponds to stiffness of the material, and marker size indicates IPF where larger markers are a higher IPF. Points with an 'x' are melting points collected from queryable sources where open circles are predicted values. a) Predicted results for original 11,000 query. b) Results filtered to remove radioactive and lanthanide compounds, and bulk and shear modulus values below 125 and 25 GPa respectively c) Additional filtering of properties with remaining values including IPF $>$ 0.4, T$_{melt}>$ 1750$^{\circ}$C, and VFE $>$ 4.5 eV/atom. d) Data shown from c) with predicted error bars. Values that have database values do not show error in respective direction. Note the slightly different scales in the filtered figures.}
\label{fig:melt_vs_vac}
\end{figure*}

Fig. \ref{fig:melt_vs_vac}(a) shows the 11,000 oxides and their respective properties. We show melting temperature and the oxygen VFE, bulk modulus is shown as the color of the symbol, and the IPF is represented by size. Fig. \ref{fig:melt_vs_vac}(b)  filters radioactive and lanthanide out, and also remove bulk and shear modulus values below 125 and 25 GPa respectively. The plot highlights common and effective protective oxides. As expected, Cr2O3, Al2O3, and SiO2 are among the top performers. However, our study reveals other oxides predicted to perform equally well or outperform them. Fig. \ref{fig:melt_vs_vac}(c) shows the final filtering of outlier properties such as low VFE, low melting point, and IPF values below 0.4 and Fig. \ref{fig:melt_vs_vac}(d) shows these final points including the uncertainties in the RF model. Data points with a cross represent materials with existing melting temperatures from WolframAlpha and empty symbols are predictions. Values without the lack of error bars in a direction indicate database collected values.

In the case of the design of refractory CCAs, HfO$_{2}$ (T$_{melt}$ = 2812$^{\circ}$C, VFE = 5.9 eV/atom) is an attractive candidate. In addition to high T$_m$ and VFE, it exhibits a CTE very close to  common RCCAs. If we were interested in considering other types of complex oxides the addition of Y as a dopant encourages the formation of Y$_{2}$Hf$_{2}$O$_{7}$, YTaO$_{4}$, Y$_{3}$Al$_{5}$O$_{12}$, or Y$_{6}$WO$_{12}$. While many of these have lower melting points (ranging from 1900-2000$^{\circ}$), they may stabilize as complex oxides between the outer scale and substrate. Each of these oxides coupled with the RCCA substrate could be engineered to form a stabilized complex oxide of one or more of these structures. 

Other notable oxides that we found with excellent properties include: MgO (T$_{melt}$ = 2852$^{\circ}$C, VFE = 6.0 eV/atom), MgAl$_{2}$O$_{4}$ (T$_{melt}$ = 2130$^{\circ}$C, VFE = 6.05 eV/atom), ZrO$_{2}$ (T$_{melt}$ = 2700$^{\circ}$C, VFE = 5.79 eV/atom), BaZrO$_{3}$ (T$_{melt}$ = 2450$^{\circ}$C, VFE = 5.63 eV/atom), ZrSiO$_{4}$ (T$_{melt}$ = 2550$^{\circ}$C, VFE = 5.70 eV/atom), and SrZrO$_{3}$ (T$_{melt}$ = 2204$^{\circ}$C, VFE = 5.6 eV/atom). We would like to stress that additional variables need to considered in the design of oxide scales, such as processability; our list is based on the properties we investigated.

It should also be noted that the uncertainty in the random forest model ranges from 20-30\% for  melting point and 10-15\% for the VFE. This indicates that additional melting temperature values during training would be desirable to narrow down material selection. However, we would like to highlight that the methods proposed in this paper are more meant to guide material selection, and not necessarily to arrive at a value of ultimate precision. These uncertainties represent those in the RF model, but in the case of melting temperature additional uncertainties can be induced due to the use of model-predicted elastic constants, this is discussed next.
 
\section{Uncertainty propagation on the melting temperature calculation}\label{UQ}

When using RF-predicted values for bulk and shear modulus as input descriptors to the melting point model, it is critical to assess how the  uncertainties in elastic constants affect the predicted T$_m$. We note that the majority of the compounds that passed the filtering steps in Section \ref{selection} had first principles elastic constant data,  one exception is BaTi$_{2}$O$_{5}$. The predicted mean melting point for this specific compound was 2144$\pm$435$^{o}$C, this was obtained with mean bulk and shear moduli. Since the elasticity models yield mean and the associated deviations, we can assess how sensitive the predicted melting point is to uncertainties in the moduli parameters. Our trained random forest models predict mean values of 142$\pm$27 and 75$\pm$16 GPa for bulk and shear modulus, respectively.

\begin{figure}[H]
\centering
\includegraphics[scale=0.40]{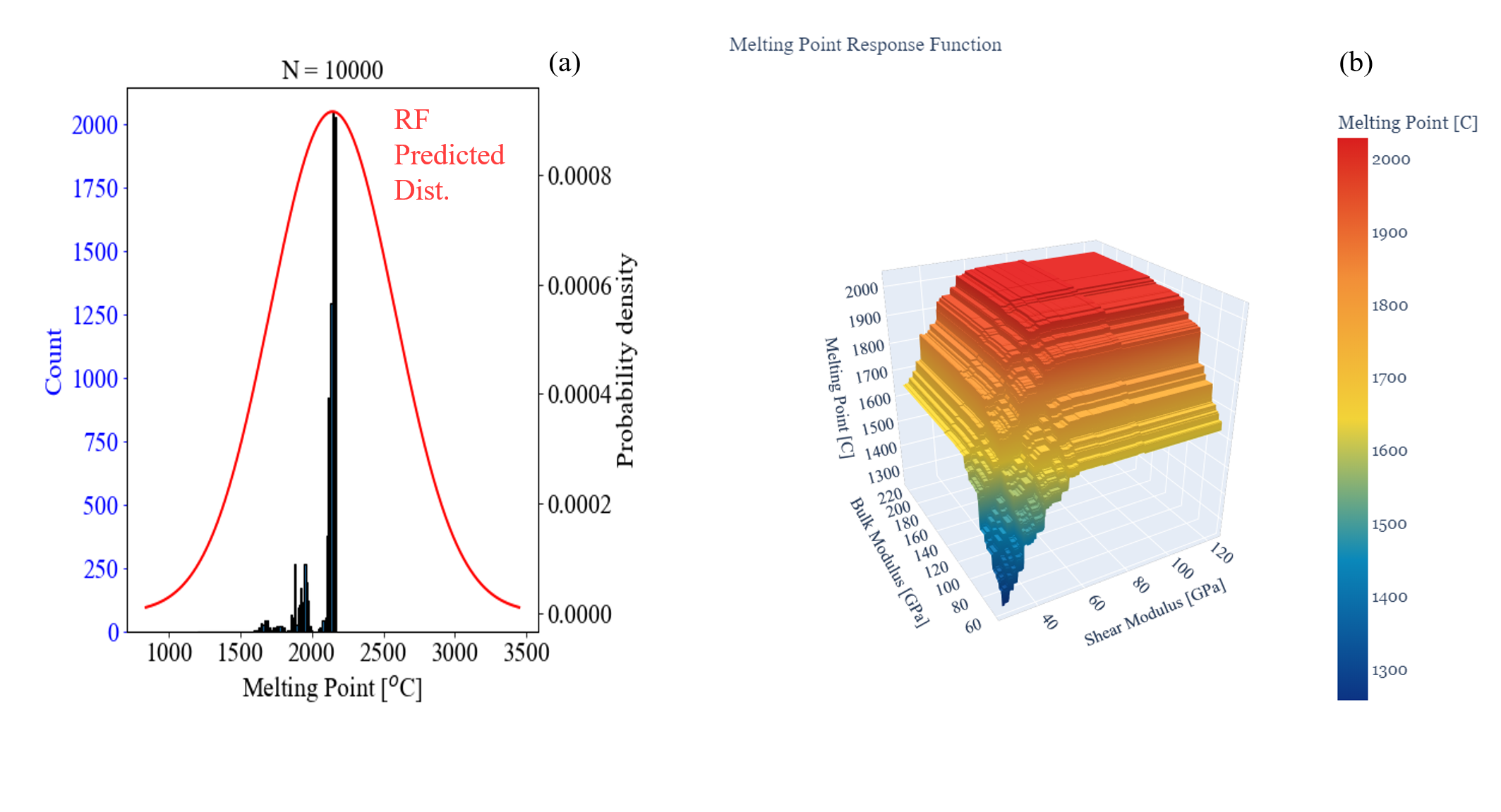}
\caption{a) Histogram results for Monte Carlo (MC) sampling of bulk and shear modulus compared to original random forest (RF) predicted distribution. b) Response surface for shear (x-axis) and bulk modulus (y-axis) with respect to melting temperature (z-axis).}
\label{fig:UQ}
\end{figure}

To propagate uncertainties in elastic constants through the melting temperature model, we use a brute force random sampling of the Gaussian distribution for each stiffness property. The resulting distribution from 10,000 samples is shown in black in Fig. \ref{fig:UQ}(a).  The predicted distribution shows a sharp peak at 2150 $^{o}$C, very close to the mean prediction, and extends towards lower values with a second peak at 1950 $^{o}$C, and a third smaller distribution centered at 1700 $^{o}$C.  The predicted RF distribution of melting temperatures with mean stiffness values is shown in red. Importantly, the uncertainties originating from the propagation of uncertainties in the stiffness are small compared with the intrinsic uncertainties in the prediction of melting temperature. This is, perhaps, not surprising since the melting temperature model has larger uncertainties that that for stiffness. The multi-peak nature of the distributions indicates large non-linearities in the T$_m$ model. To assess this, we plot the melting temperature as a function of shear and bulk modulus in Fig. \ref{fig:UQ}(b) with all other paramters fixed to those of BaTi$_{2}$O$_{5}$. We find that melting temperature drops quite significantly for low values of shear and bulk moduli. This is not surprising given the positive correlation between stiffness and melting temperature, but such extrapolations using machine learning models should be done with care.

\section{Summary and outlook}

We showed that by leveraging queryable open repositories and the use of machine learning tools with infused physics one can greatly expand the information available for materials design or selection. Our specific goal was to find oxides for high temperature applications with high melting temperature, high oxygen vacancy formation energy (to minimize O transport) with coefficient of thermal expansion and stiffness as secondary variables. Quantifying uncertainty still remains a key aspect, and continued research into assessing uncertainty is more robust machine learning models such as neural networks will be critical.

Machine learning models with physics insight built in via feature engineering and surrogate properties enables us to take sparse existing data and fill-in gaps in knowledge. Through these methods we were able to expand an initial query of 162 melting points and push our predictive capability into a space two orders of magnitude higher. Of the oxide space explored there are many candidates that compete with current industrial oxides such as Al$_{2}$O$_{3}$ and Cr$_{2}$O$_{3}$ with respect to melting point, VFE, IPF, and stiffness.

We find that through this rapid queries and machine learning prediction coupling that suite of RCCA containing oxides including:  HfO$_{2}$,  Y$_{2}$Hf$_{2}$O$_{7}$, YTaO$_{4}$, Y$_{3}$Al$_{5}$O$_{12}$, Y$_{6}$WO$_{12}$, MgO, MgAl$_{2}$O$_{4}$, ZrO$_{2}$ , BaZrO$_{3}$ , ZrSiO$_{4}$ , and SrZrO$_{3}$ could be likely candidates for stabilizing oxide formation in high temperature alloy applications. 

The most critical aspect of this oxide search has been the accessibility and wealth of data found on materials informatics platforms. Contribution to these databases remains key, and we intend to supplement the Materials Project Database with more calculations for elasticity from first principles, and the curated datasets from this project will be made available on Citrination for public use. Additional extensions of queried data to supplement our models from the previously mentioned databases is a continued area of work. 

The models built and developed in this paper can be accessed through the nanoHUB tool High Temperature Oxide Property Explorer \cite{33792}. Final curated data can be downloaded, and models can be modified at the leisure of the user.

\section{Acknowledgements}
Insightful discussions with Prof. M. Titus and K. Sandhage of Purdue University are gratefully acknowledged. This effort was supported by the US National Science Foundation, DMREF program under contract number 1922316-DMR. We would like to acknowledge the use and support of computational resources from nanoHUB and Purdue University RCAC.


\bibliography{Library.bib}

\end{document}